# Single Pair of Charge-two Weyl Fermions in Chiral Boron Allotropes

Hui-Jing Zheng[1], Yan Gao[1]*, Yanfeng Ge[1], Yong Liu[1], and Zhong-Yi Lu[2,3]

[1]*State Key Laboratory of Metastable Materials Science and Technology & Hebei Key Laboratory of Microstructural Material Physics, School of Science, Yanshan University, Qinhuangdao 066004, China*

[2]*School of Physics and Beijing Key Laboratory of Opto-electronic Functional Materials & Micro-nano Devices, Renmin University of China, Beijing 100872, China*

[3]*Key Laboratory of Quantum State Construction and Manipulation (Ministry of Education), Renmin University of China, Beijing 100872, China*

**The realization of a minimal Weyl semimetal (WSM) hosting a single pair of Weyl points (WPs) has thus far been restricted to magnetic systems, since time-reversal symmetry generally enforces a minimum of four WPs in nonmagnetic materials. Here, combining first-principles calculations with symmetry analysis, we identify two stable boron allotropes, chiral HDSBC-$B_{20}$ and cage-like CR-$B_{12}$, as the first nonmagnetic electronic materials realizing a single pair of WPs in the spinless regime. We show that the interplay between time-reversal symmetry and crystallographic rotation symmetry ($C_4$ or $C_3$) stabilizes exactly one pair of charge-2 WPs pinned at time-reversal-invariant momenta, thereby circumventing the conventional node-quartet constraint. These double-WPs exhibit linear dispersion along the rotation axis and quadratic dispersion in the perpendicular plane. In HDSBC-$B_{20}$, the sign of the topological charge is directly correlated with structural chirality. Both materials host exceptionally long double Fermi arcs spanning the surface Brillouin zone, providing experimentally accessible signatures. Our findings establish nonmagnetic material platforms for minimal double-Weyl fermions and broaden the landscape of unconventional WSMs.**

*Corresponding authors: yangao9419@ysu.edu.cn

**Introduction**

The discovery of Weyl semimetals (WSMs) has opened a new frontier in condensed matter physics[1-4], providing a solid-state realization of Weyl fermions[5-7] and giving rise to a host of topological phenomena, including the chiral anomaly[8,9], topological Fermi-arc surface states[1-3], quantum anomalous Hall effect[10], and negative magnetoresistance effect[11,12]. These phenomena originate from bulk Weyl points (WPs)[5], which act as monopoles of Berry curvature in momentum space and are characterized by quantized topological charges, or Chern numbers $C$. Conventional WPs with $|C| = 1$ exhibit linear band dispersion along all momentum directions, whereas unconventional multi-Weyl points[13] with higher chiral charges ($|C| = 2,3,4$) display anisotropic dispersions in which two distinct orders of momentum dependence, among linear, quadratic, and cubic, are combined[14]. Such higher-charge WPs are necessarily stabilized by crystalline rotational symmetries[13-16] and are expected to host enhanced topological responses, including longer Fermi arc surface states[17], quantum criticality and phase transition[18], larger quantized circular photogalvanic effect[19], and possible deviations from Fermi-liquid behavior[20]. A central objective in Weyl physics is the realization of minimal WSMs hosting the fewest possible WPs, ideally a single pair. Such minimal configurations avoid the complications introduced by multiple WPs and provide the conceptually cleanest setting for establishing bulk-boundary correspondence.

However, in nonmagnetic systems this goal is strongly constrained by the Nielsen–Ninomiya no-go theorem[21,22]. Time-reversal symmetry ($\mathcal{T}$) enforces the appearance of WPs with identical chirality at $k$ and -$k$, thereby requiring at least four WPs (two pairs) to satisfy overall topological charge neutrality[23-28]. Consequently, circumventing this "four-node" barrier to realize a minimal WSM hosting only a single Weyl-point pair has emerged as a key frontier in the field. Although theoretical strategies have been proposed to circumvent this limitation by pinning WPs at time-reversal-invariant momenta (TRIMs), a mechanism permitting a single-pair $|C| = 2/4$ WPs in the spinless systems[29], experimental and theoretical material realizations have thus far been largely restricted to bosonic systems[30, 31]. Magnetic systems can bypass this constraint by breaking $\mathcal{T}$ to host a single Weyl pair, as exemplified by β-$V_2OPO_4$[32], $K_2Mn_3(AsO_4)_3$[33], EuAgP[34], $EuCd_2As_2$[35] and $MnSn_2Sb_2Te_6$-B[36]. While these works have significantly advanced the understanding of minimal Weyl configurations, the resulting WPs are inextricably tied to magnetic ordering, which is often realized only under specific conditions such as low temperatures or finely tuned magnetic structures. In the nonmagnetic electronic realm, however, realizing a pristine minimal Weyl state remains an elusive goal, frequently obscured by strong spin–orbit coupling (SOC) and the entanglement of trivial bands near the Fermi level[37, 38].

In this work, combining first-principles calculations with symmetry analysis, we overcome this challenge by identifying two boron allotropes, chiral HDSBC-$B_{20}$ and cage-structured CR-$B_{12}$, that realize the long-sought single pair of double-Weyl fermions in a nonmagnetic electronic environment. Both structures are demonstrated to be dynamically, thermally, and mechanically stable, with CR-$B_{12}$ notably exhibiting a formation energy lower than many topological boron phases and even experimentally synthesized $\alpha$-Ga boron. Crucially, the electronic structure near the Fermi level is pristine: the valence and conduction bands intersect exclusively at two distinct TRIMs, forming a single pair of $|C|=2$ WPs protected by $C_4/C_3$ rotation and $\mathcal{T}$, free from any coexisting trivial Fermi pockets. In chiral HDSBC-$B_{20}$, the sign of the Weyl-point topological charge is strictly dictated by the structural handedness, establishing a direct link between real-space chirality and momentum-space topology. The large separation of the WPs, combined with their double charge, generates "maximal" double Fermi arcs that traverse the surface Brillouin zone (BZ), offering unambiguous experimental fingerprints. Our findings establish an ideal platform for investigating minimal $|C|=2$ Weyl fermions and open a new paradigm for realizing unconventional Weyl physics in light-element, nonmagnetic systems.

**Results**

The structural versatility of boron allows for the formation of complex three-dimensional (3D) networks. Here, we identify two distinct metastable boron allotropes. The first, HDSBC-$B_{20}$, crystallizes in the chiral tetragonal space groups $P4_3$ (No. 78) for left-handed *l*-HDSBC-$B_{20}$ [Fig. 1(a)] and $P4_1$ (No. 76) for right-handed *r*-HDSBC-$B_{20}$ [Fig. 1(b)]. These enantiomers, related by spatial inversion, are characterized by homochiral boron chains of distinct widths: single-atom-width spiraling boron chains (SBC) [Fig. 1(c)] and double-atom-width helical chains (HD) [Fig. 1(d)] are interconnected via bridging atoms (green) to form a stable 3D network. Each primitive unit cell contains 20 boron atoms [Fig. 1(e)], hence the designation HDSBC-$B_{20}$ (Helical Double & Single Boron Chain). The second phase, CR-$B_{12}$, adopts the rhombohedral space group $R32$ (No. 155). Its primitive cell comprises a new type of $B_{12}$ icosahedral cages [Fig. 1(j)] forming a covalent cage-based network [Figs. 1(g)-1(h)], distinct from the conventional $\alpha$-$B_{12}$ motif [Fig. 1(k)].

The optimized structural parameters are summarized in Table SI [see Supplemental Material (SM)]. Remarkably, CR-$B_{12}$ exhibits a formation energy of -6.52 eV/atom, which, while slightly higher than the near-ground-state $\alpha$-$B_{12}$[39], is energetically lower than that of many proposed topological boron phases and even lower (by 0.11 eV/atom) than the experimentally synthesized $\alpha$-Ga phase[40]. Comprehensive stability analysis further confirms their viability: (i) Phonon dispersions show no imaginary frequencies across the BZ for CR-$B_{12}$ [Fig. 2(c)], and only exhibit negligible soft modes near Γ for HDSBC-$B_{20}$ (likely of numerical origin), collectively confirming the dynamical stability of both phases. (ii) *Ab initio* molecular dynamics simulations confirm the room-temperature thermodynamic stability of HDSBC-$B_{20}$, with no structural bond breaking observed at 300 K [Fig. 2(b)], while CR-$B_{12}$ remains structurally stable up to 900 K [Fig. 2(d)]. (iii) Mechanical stability is rigorously verified, as the calculated elastic stiffness constants $C_{ij}$ fully satisfy the Born stability criteria[41] for both tetragonal and trigonal symmetries [see Table SII in the SM].

We first examine the electronic structure of the chiral boron allotrope HDSBC-B$_{20}$. Because boron is a light element with intrinsically weak SOC, SOC effects are negligible, and the system can be accurately described within a spinless framework ($\mathcal{T}^2 = 1$). The band structure and projected density of states (PDOS) for $l$-HDSBC-B$_{20}$ are presented in Figs. 3(a) and 3(b), respectively. The PDOS exhibits a pronounced minimum near the Fermi level $E_F$, characteristic of semimetallic behavior. Inspection of the bulk bands reveals that the valence (band 30) and conduction (band 31) states touch exclusively at two isolated TRIMs in the BZ: W$_1$ at Γ ($E - E_F = 0.155$ eV) and W$_2$ at M ($E - E_F = -0.052$ eV). A comprehensive scan of the BZ confirms that a finite gap persists everywhere else between these two bands, strictly satisfying the minimal node count mandated by the Nielsen-Ninomiya theorem for a time-reversal invariant system.

Symmetry analysis of the Bloch states confirms the topological nature of these crossings. The irreducible representations at Γ and M correspond to the doubly degenerate $\Gamma_3\Gamma_4$ and $M_3M_4$ states of space group $P4_3$, respectively [Fig. 3(a)], identifying each node as a protected degeneracy. The band dispersions in the vicinity of W$_1$ and W$_2$, plotted in the $k_x$–$k_y$ plane [Figs. 3(c) and 3(d)], clearly exhibit quadratic dispersion in the transverse plane, while Fig. 3(a) reveals linear dispersion along the $k_z$ (Γ-Z and M-A) direction. Such a linear-quadratic dispersion is the hallmark of double-Weyl fermions with topological charge $|C| = 2$.

To elucidate the symmetry protection mechanism, we construct a low-energy $\boldsymbol{k} \cdot \boldsymbol{p}$ effective model. At Γ point (W$_1$), the little group is $C_4$ within space groups $P4_3$ (No. 78), generated by fourfold rotation $C_4$ and time-reversal symmetry $\mathcal{T}$. In the basis of the $\Gamma_3\Gamma_4$ doublet, the matrix representations of symmetry operators are represented as $D(C_{4,001}^+) = i\sigma_z$ and $D(\mathcal{T}) = \sigma_x K$, where $\sigma_i$ ($i = x, y, z$) are Pauli matrixes and $K$ is the complex-conjugate operator. These generators impose the following constraints on the Hamiltonian:

$$D(g)H(\boldsymbol{k})D^{-1}(g) = H(g\boldsymbol{k}), \tag{1}$$

where $g$ represents the generating elements of $C_{4,001}^+$ and $\mathcal{T}$. Hence, the two band $\boldsymbol{k} \cdot \boldsymbol{p}$ effective Hamiltonian expanded up to $k^2$ order is derived as

$$H_{W_1}(\boldsymbol{k}) = \varepsilon(\boldsymbol{k}) + [(\alpha k_+^2 + \beta k_-^2)\sigma_+ + \text{H.c.}] + m k_z \sigma_z, \tag{2}$$

where $\varepsilon(\boldsymbol{k}) = \epsilon_0 + \epsilon_1(k_x^2 + k_y^2) + \epsilon_2 k_z^2$. $\epsilon_i$ ($i = 0,1,2$) and $m$ are real parameters, while $\alpha$ and $\beta$ are complex parameters, all determined by fitting to the DFT band structure of HDSBC-B$_{20}$. This model explicitly confirms that $W_1$ at $\Gamma$ is a double-Weyl point with linear dispersion along $k_z$ and quadratic dispersion in the $k_x$–$k_y$ plane. An analogous symmetry analysis at the M point ($W_2$) yields an identical effective Hamiltonian, establishing the double-Weyl nature of $W_2$ as well. The optimized model parameters reproduce the DFT bands with excellent agreement [see Figs. S1(a) and S1(b) in the SM], validating the physical reliability of the effective model.

The topological charges of $W_1$ and $W_2$ are unambiguously determined by integrating the Berry curvature over a closed surface enclosing each node. We find $C = -2$ for $W_1$ and $C = +2$ for $W_2$ [Fig. 3(e)], satisfying the net topological charge neutrality. The Berry curvature distribution in the $k_z = 0$ plane [Fig. 3(f)] further visualizes $W_2$ as a Berry-curvature source and $W_1$ as a sink, in full consistency with their respective chiral charges.

Upon projecting the bulk WPs onto the (001) surface [Fig. 1(f)], we observe prominent Fermi-arc surface states crossing the Fermi level [Fig. 3(g)]. These arcs directly connect the projected WPs with opposite topological charges, providing a clear manifestation of the bulk–boundary correspondence. At the energy slice $E = -52$ meV, corresponding to the energy of $W_2$, Fig. 3(h) reveals that the surface states consist of extended double Fermi arcs across the surface BZ, a direct consequence of the minimal single-pair double-WPs configuration. Calculations for the (100) surface [Fig. S2 in the SM] yield similarly long and extended double Fermi arcs. Together, both surfaces exhibit robust and distinctive signatures of a single pair of double-WPs, providing distinctive signals for future experimental detection.

For the right-handed counterpart $r$-HDSBC-B$_{20}$ [Figs. 3(i-p), the overall electronic structure remains qualitatively identical to that of the left-handed $l$-HDSBC-B$_{20}$ [Figs. 3(i-l)]. Importantly, however, the chiralities of the WPs are reversed [Figs. 3(m-n)], in direct correspondence with the inversion of the crystal handedness. This one-to-one mapping between structural chirality and Weyl-point chirality provides a clear physical mechanism for controlling topological charge through crystal geometry, suggesting potential applications in chirality-dependent transport and optical responses. Consistent with this picture, $r$-HDSBC-B$_{20}$ also exhibits extended double Fermi arcs on both the (001) and (100) surfaces [Figs. 3(o-p) and Fig. S3 in the SM].

We next turn to the electronic structure and topological properties of CR-B$_{12}$, which exhibits a closely related but symmetry-distinct realization of single-pair double-Weyl fermions. Figures 4(a) and 4(b) show the band structure and PDOS, respectively. A pronounced minimum in the PDOS near the intrinsic Fermi level again indicates a semimetallic character. The valence-band maximum and conduction-band minimum (18th and 19th bands) intersect only at two points in the entire BZ: $W_1$ at Γ and $W_2$ at the T point. Irreducible-representation analysis reveals that $W_1$ and $W_2$ transform as the two-dimensional (2D) representations $Γ_3$ and $T_3$, respectively, implying that both are doubly degenerate WPs. Energetically, $W_1$ is located 0.902 eV above $E_F$, while $W_2$ lies 0.133 eV below $E_F$ [see Table SIII in the SM]. The PDOS indicates that the low-energy states near both WPs are dominated by boron $p_z$ orbitals. As shown in Figs. 4(c) and 4(d), both $W_1$ and $W_2$ display quadratic dispersion in the $k_x$–$k_y$ plane and linear dispersion along the threefold rotation axis (Γ–T direction), a characteristic signature of $|C| = 2$ WPs.

The effective Hamiltonian at Γ for space group $R32$ ($D_3$ point group) is constrained by $C_{3,001}^+$, $C_{2,100}$, and time-reversal symmetry $\mathcal{T}$. Under the $Γ_3$ basis, the corresponding matrix representations are

$$D(C_{3,001}^+) = \begin{bmatrix} e^{\frac{i2\pi}{3}} & 0 \\ 0 & e^{-\frac{i2\pi}{3}} \end{bmatrix}, \quad D(C_{2,100}) = \begin{bmatrix} 0 & e^{-\frac{i2\pi}{3}} \\ e^{\frac{i2\pi}{3}} & 0 \end{bmatrix}, \quad D(T) = \sigma_x K. \quad (3)$$

These generators impose symmetry constraints on the Hamiltonian through Eq. (1). Expanding to leading order yields

$$H_{W_1}(\bm{k}) = \varepsilon(\bm{k}) + [(\alpha k_+^2 + \beta k_- k_z)\sigma_+ + \text{H.c.}] + mk_z\sigma_z, \quad (4)$$

where $\varepsilon(\bm{k}) = \epsilon_0 + \epsilon_1(k_x^2 + k_y^2) + \epsilon_2 k_z^2$. Fitting to the DFT bands confirms that $W_1$ at $\Gamma$ is a double-Weyl point with quadratic in-plane and linear out-of-plane dispersion [Fig. S1(c)]. An analogous construction at the T point yields the same Hamiltonian form, establishing the double-Weyl nature of $W_2$ [Fig. S1(d)].

The chiral charges are determined by Berry-curvature integration, giving $C = +2$ for $W_1$ and $C = -2$ for $W_2$ [Fig. 4(e)]. The corresponding Berry-curvature distribution [Fig. 4(f)] identifies $W_1$ as a source and $W_2$ as a sink. Surface-state calculations reveal clear double Fermi arcs on the (001) surface at $E - E_F = 80$ meV [Figs. 4(g) and 4(h)], connecting the projections of $W_1$ and $W_2$. Similar features are found on the (110) surface [Fig. S4 in the SM], demonstrating that these features are not surface-specific but instead represent a robust consequence of the bulk topology.

**Discussion**

A crucial aspect of the present study concerns the role and limitation of SOC. From a symmetry standpoint, it has been rigorously shown that a minimal Weyl configuration consisting of a single pair of charge-2 Weyl points in a nonmagnetic system is strictly allowed only in the spinless regime ($\mathcal{T}^2 = +1$), whereas the inclusion of SOC generally renders such a configuration symmetry-forbidden due to Kramers degeneracy and double-valued representations[29]. In this sense, the spinless description adopted here is not merely a computational approximation but a fundamental requirement imposed by symmetry. This constraint also explains why previous searches for minimal WSMs in nonmagnetic electronic materials, largely focused on heavy-element systems with strong SOC, have so far been unsuccessful. Motivated by this consideration, our work targets light-element boron allotropes, whose intrinsically weak SOC enables the realization of this long-sought Weyl phase in an electronic setting. Explicit SOC calculations show that SOC opens only extremely small gaps near the WPs of HDSBC-$B_{20}$ and CR-$B_{12}$, ranging from 0.08 meV to 1.19 meV (corresponding to energy scales of approximately 0.93–13.81 K) [see Fig. S5], which are negligible under typical experimental conditions. Therefore, the predicted single-pair double-Weyl fermions and their associated surface states remain physically relevant over a broad temperature range. These results establish light-element nonmagnetic systems with negligible SOC as an essential materials platform for realizing the simplest possible electronic WSM.

A symmetry-based mechanism for realizing a single pair of charge-2 Weyl points in spinless systems was previously proposed by Wang *et al.* [29]. Our work advances this idea by demonstrating its realization in realistic nonmagnetic electronic materials and by revealing their concrete electronic and surface-state signatures. The identification of single-pair double-Weyl fermions in *l*/*r*-HDSBC-$B_{20}$ and CR-$B_{12}$ represents a substantial advance, establishing them as the first nonmagnetic electronic materials hosting this minimal $|C| = 2$ Weyl configuration. Previously reported $|C| = 2$ systems, such as $(TaSe_4)_2I$[42], $CaGe_2$[43], and $SrSi_2$[44], contain multiple Weyl-point pairs, complicating the low-energy physics through inter-node coupling. While single-pair WPs have been realized in bosonic platforms such as chiral photonic and phononic crystals[30, 31], these systems inherently lack fermionic (electronic) degrees of freedom. Consequently, they cannot host electronic charge transport, enable electronic device functionalities, or couple directly to electromagnetic fields in the manner of WSMs. Magnetic WSMs, such as β-$V2OPO4$[32], $K_2Mn_3(AsO_4)_3$[33], $EuAgP$[34], $EuCd_2As_2$[35] and $MnSn_2Sb_2Te_6$-B[36], achieve single-pair WPs configurations only under specific magnetic ordering conditions, typically at low temperatures.

In contrast, *l*/*r*-HDSBC-$B_{20}$ and CR-$B_{12}$ combine several decisive advantages: they are nonmagnetic and thus operable at room temperature; they are electronic systems supporting charge transport, optical responses, and field-effect tunability; their light-element composition leads to negligible SOC, enabling the realization of a single-pair WPs in a spinless setting; their low-energy electronic structures are exceptionally clean, involving only two bands near $E_F$; the large momentum-space separation between the WPs produces extraordinarily long double Fermi arcs readily accessible to ARPES and STM; in HDSBC-$B_{20}$, the Weyl-point chirality is intrinsically tied to the crystal handedness, opening avenues for chirality-dependent optoelectronic effects; and CR-$B_{12}$ exhibits formation energies comparable to experimentally synthesized boron allotropes, suggesting realistic synthetic feasibility.

In summary, we have theoretically identified two boron allotropes, HDSBC-$B_{20}$ and CR-$B_{12}$, as the first nonmagnetic electronic materials hosting a minimal single pair of double-Weyl fermions. Both systems host exactly one pair of WPs carrying topological charge $|C| = 2$ Weyl configurations protected by $C_3/C_4$ rotational and time-reversal symmetries and give rise to exceptionally long double Fermi arcs extending across the surface BZ. Our work not only enriches the physics of boron materials but also shifts the focus from magnetic or bosonic systems to nonmagnetic light-element electronic platforms in the search for the minimal Weyl semimetal.

**Methods**

**Computational methods:** The electronic structures of HDSBC-$B_{20}$ and CR-$B_{12}$ were investigated by using the projector augmented wave (PAW) method [45] as implemented in the VASP package [46] in the framework of density functional theory (DFT). The exchange-correlation potential was treated using the generalized gradient approximation (GGA) of Perdew-Burke-Ernzerhof (PBE) type [47]. The plane-wave kinetic energy cutoff was set to 550 eV. Brillouin zone (BZ) integration utilized Γ-centered $k$-point meshes [48] of $4 \times 4 \times 6$ for HDSBC-$B_{20}$ and $8 \times 8 \times 8$ for CR-$B_{12}$. Convergence criteria for energy and forces were set to $10^{-6}$ eV and $10^{-3}$ eV/Å, respectively. Phonon dispersions were computed using the finite displacement method implemented in the PHONOPY package [49] on $3 \times 3 \times 3$ (HDSBC-$B_{20}$) and $2 \times 2 \times 2$ (CR-$B_{12}$) supercells, respectively. Thermal stability was verified via *ab-initio* molecular dynamics (AIMD) simulations in a canonical ensemble using a Nose-Hoover thermostat [50]. Elastic stiffness constants $C_{ij}$ derived using the energy-strain method [51]. Topological properties were analyzed using WannierTools codes[52] based on effective tight-binding models derived from maximally localized Wannier functions via Wannier90 [53].

**Data availability**

The data supporting the findings are displayed in the main text and the Supplementary Information. All raw data are available from the corresponding authors upon request.

## Acknowledgements:

We wish to thank Weikang Wu and Shengyuan A. Yang for helpful discussions. This work was supported by the National Natural Science Foundation of China (Grants No. 12304202), Hebei Natural Science Foundation (Grant No. A2023203007), Science Research Project of Hebei Education Department (Grant No. BJK2024085), and Cultivation Project for Basic Research and Innovation of Yanshan University (No. 2022LGZD001).

## Author contributions:

Hui-Jing Zheng proposed the HDSBC-$B_{20}$ and CR-$B_{12}$ structures and performed the corresponding DFT calculations. Yan Gao conceived the original ideas. Yanfeng Ge, Yong Liu, and Zhong-Yi Lu analyzed the results. Yan Gao writing-reviewing, conceptualization, supervision, and project administration. All authors discussed the results and commented on the manuscript at all stages.

## Competing interests

The authors declare no competing financial interests.

# Figures and Tables

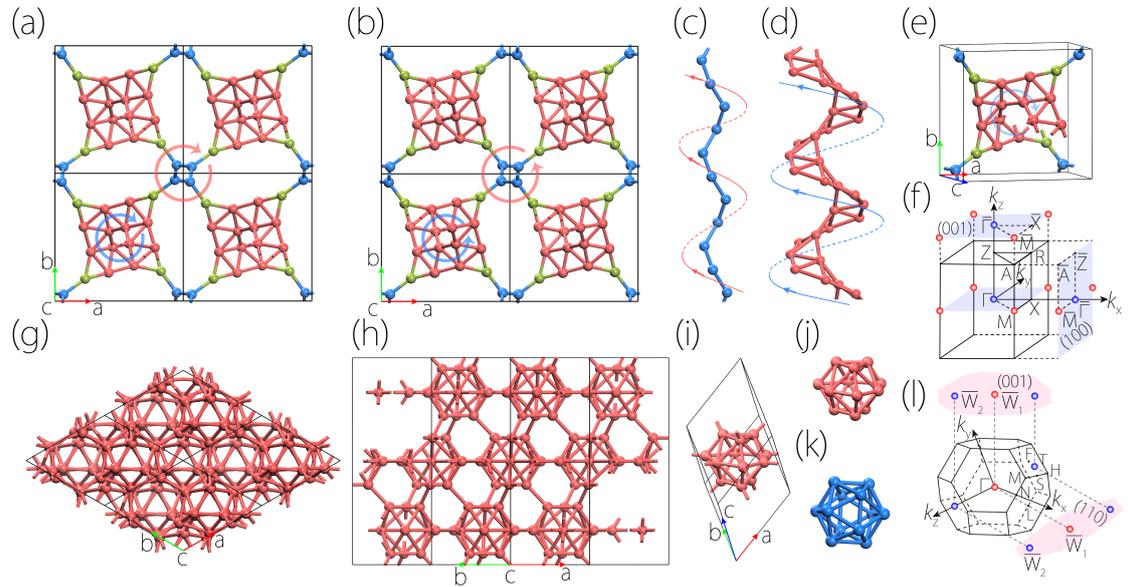

**Figure 1 | Crystal structures and Brillouin zone (BZ) of HDSBC-$B_{20}$ with two enantiomeric left/right-handed structures and CR-$B_{12}$.** (a) Top view of left-handed *l*-HDSBC-$B_{20}$. (b) Top view of right-handed *r*-HDSBC-$B_{20}$. (c) Single-atom-wide helical boron chain (blue) of *l*-HDSBC-$B_{20}$. (d) Double-atom-wide helical boron chain (red) of *l*-HDSBC-$B_{20}$. (e) Perspective view of the unit cell of *l*-HDSBC-$B_{20}$. (f) Distribution of Weyl points of *l*-HDSBC-$B_{20}$ in the 3D BZ showing $C = -2$ and $C = +2$ at Γ and M positions, respectively, and their projections on the (001) and (100) surfaces. (g) Top and (h) side views of the $2 \times 2 \times 1$ conventional cell of CR-$B_{12}$. (i) Primitive cell of CR-$B_{12}$. (j) Basic structural unit of CR-$B_{12}$ (red). (k) Basic structural unit of α-$B_{12}$ (blue). (l) Distribution of Weyl points in the bulk BZ of CR-$B_{12}$ and their projections on the (001) and (110) surfaces.

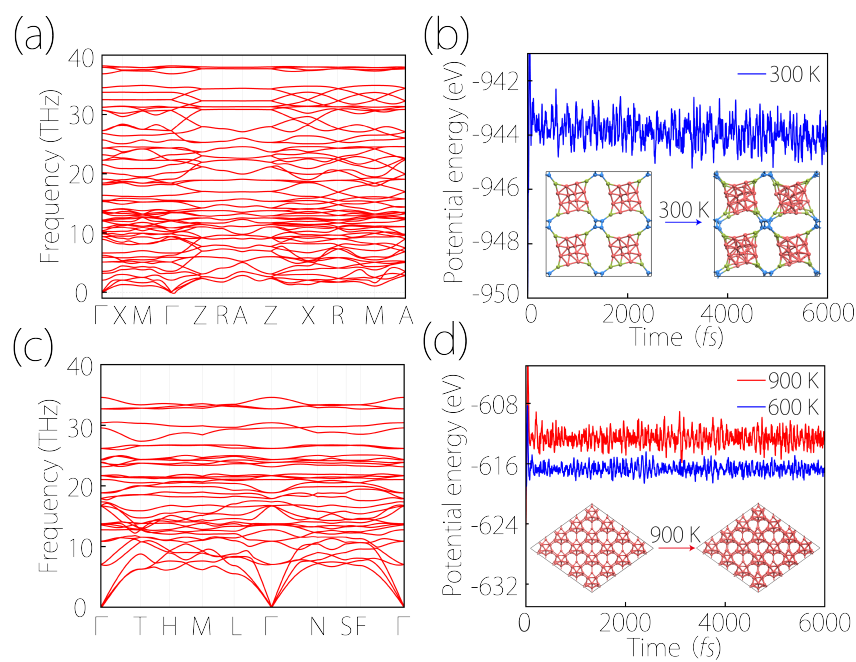

**Figure 2 | Dynamic and thermal stability of HDSBC-$B_{20}$ and CR-$B_{12}$. (a) Phonon dispersion of HDSBC-$B_{20}$ along the entire high-symmetry path.** (b) Total potential energy fluctuation in *ab initio* molecular dynamics (AIMD) simulations of HDSBC-$B_{20}$ at 300 K. The inset shows the equilibrium atomic configuration after 6000 fs relaxation at 300 K. (c) Phonon dispersion of CR-$B_{12}$ along the entire high-symmetry path. (d) Total potential energy fluctuations in AIMD simulations of CR-$B_{12}$ at 900 K and 600 K, respectively. The inset shows the equilibrium atomic configuration after 6000 fs relaxation at 900 K.

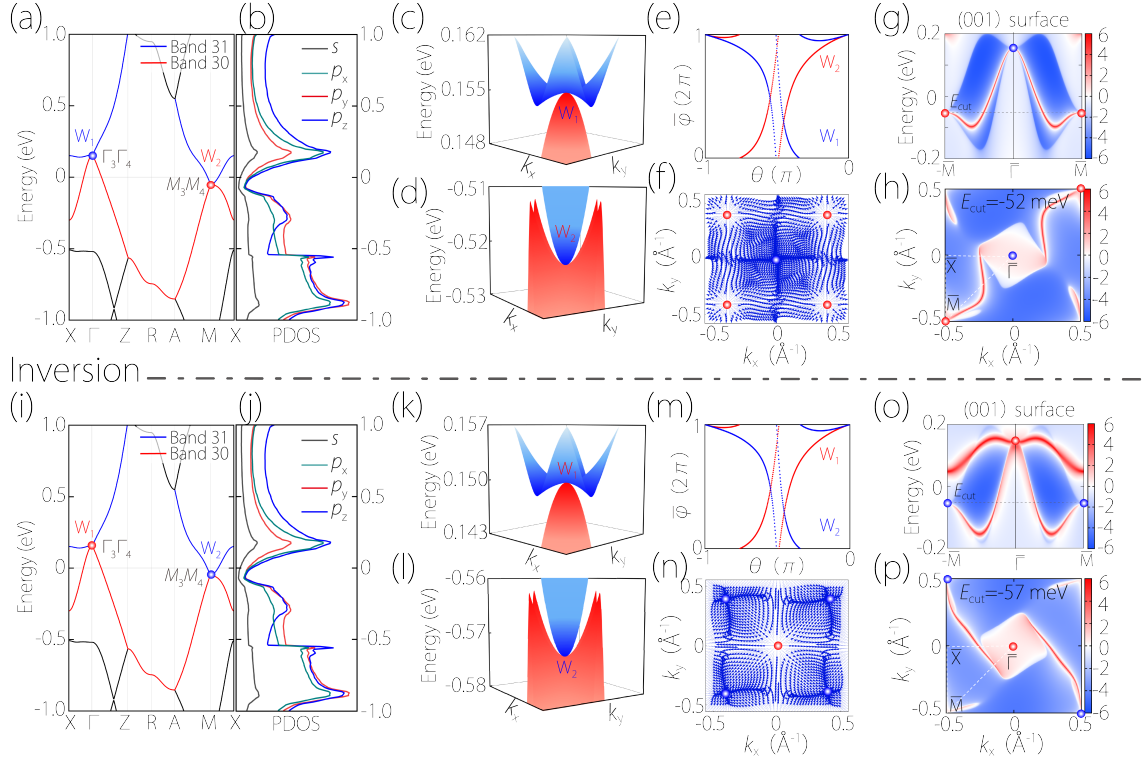

**Figure 3 | Electronic and topological properties of the chiral boron allotropes *l*-HDSBC-B$_{20}$ and *r*-HDSBC-B$_{20}$.** (a-h) Calculated electronic and topological properties for the left-handed *l*-HDSBC-B$_{20}$. (a) Band structure of *l*-HDSBC-B$_{20}$ along high-symmetry paths. Weyl points W$_1$ (blue dot) and W$_2$ (red dot) are formed by the crossing of the highest occupied 30th band and lowest unoccupied 31st band at time-reversal invariant points Γ and M. (b) Projected density of states (PDOS) of *l*-HDSBC-B$_{20}$. (c-d) 3D band structures showing quadratic dispersion characteristics of W$_1$ and W$_2$ bands in the $k_x$-$k_y$ plane, respectively. (e) Evolution of Wannier charge centers (WCCs) on spheres enclosing W$_1$ and W$_2$, showing topological charges of -2 and +2, respectively. (f) Berry curvature distribution in the $k_z = 0$ plane of *l*-HDSBC-B$_{20}$. (g) Calculated surface band structure of the (001) surface. The dashed line indicates the energy corresponding to Fig. 3(h). (h) Fermi arcs at the energy slice $E = -52$ meV (i.e., the energy of W$_2$) on the (001) surface. (i-p) The right-handed *r*-HDSBC-B$_{20}$ structure related by space inversion exhibits similar electronic and topological properties to the left-handed *l*-HDSBC-B$_{20}$, except for the reversed chirality of Weyl points.

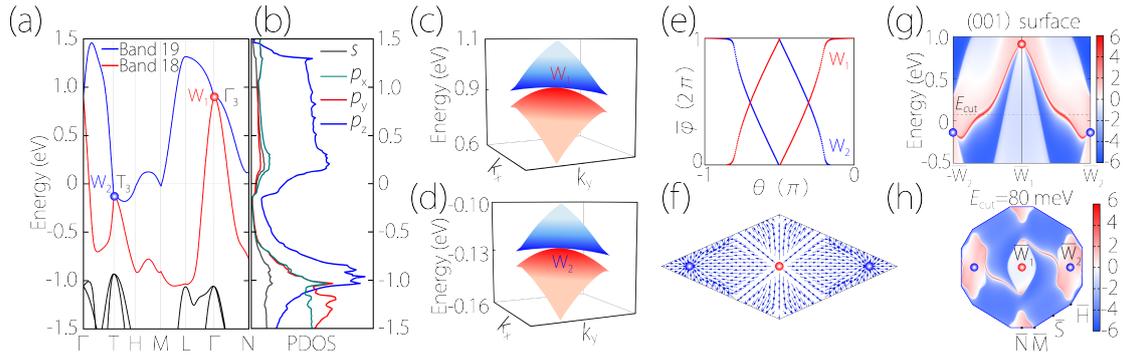

**Figure 4 | Electronic and topological properties of CR-B$_{12}$.** (a) Band structure of CR-B$_{12}$ along high-symmetry paths. Weyl points W$_1$ (red dot) and W$_2$ (blue dot) are formed by the crossing of the highest occupied 18th band and lowest unoccupied 19th band at time-reversal invariant points Γ and T. (b) Projected density of states (PDOS) of CR-B$_{12}$. (c-d) 3D band structures showing quadratic dispersion characteristics of W$_1$ and W$_2$ bands in the $k_x$–$k_y$ plane, respectively. (e) Evolution of Wannier charge centers on spheres enclosing W$_1$ and W$_2$, generating topological charges of +2 and -2, respectively. (f) Berry curvature distribution in the plane containing W$_1$ and W$_2$, confirming W$_1$ as a source and W$_2$ as a sink. (g) Calculated surface band structure of the (001) surface. The dashed line indicates the energy level corresponding to Fig. 4(h). (h) Fermi arcs at the energy slice $E = 80$ meV on the (001) surface.